\documentstyle[12pt,amsfonts]{article}
\def\iint{\int \int}

\begin{document}

\author{B.G. Konopelchenko\thanks{%
E-mail: konopel@le.infn.it} \\
Dipartimento di Fisica, Universita' di Lecce, \\
and \\
Sezione INFN, 73100, Lecce, Italy,\\
and \\
IINS, Novosibirsk Branch, Russia}
\title{On $\overline{\partial }$-problem and integrable equations}
\date{}
\maketitle

\begin{abstract}
Using the $\overline{\partial }$-problem and dual $\overline{\partial }$%
-problem , we derive bilinear relations which allows us to construct
integrable hierarchies in different parametrizations, their
Darboux-B\"{a}cklund transformations and to analyze constraints for them in
a very simple way. Scalar KP, BKP and CKP hierarchies are considered as
examples.
\end{abstract}

There are different methods to construct integrable equations and to analyze
their properties (see \textit{e.g.} [\ref{zmnp}-\ref{segal}]). The $%
\overline{\partial }$-dressing method proposed in [\ref{dbar}] is, perhaps,
one of the most effective of them. Recently, it has been applied
successfully to several important problems in soliton theory (see \textit{%
e.g.} [\ref{bogman}-\ref{doliwa}]).

In this letter we would like to attract an attention to one more profitable
aspect of the $\overline{\partial }$-dressing method. Namely, starting with $%
\overline{\partial }$-problem and dual $\overline{\partial }$-problem, we
derive two important bilinear relations for the so-called
Cauchy-Baker-Akhiezer (CBA) functions associated with different kernels $R$
of the $\overline{\partial }$-problem. These relations provide us simple
variational relations for CBA functions and $\overline{\partial }$-kernel $R$%
. In a simple unified manner, they generate integrable hierarchies in
different parametrizations and corresponding bilinear Hirota identities.
These bilinear relations are convenient also for analysis of different
constraints. It is shown how scalar BKP and CKP hierarchies arise within
such an approach. We demonstrate also that pole type parametrization of
evolutions leads to the continuous analogs of the Darboux system.

The $\overline{\partial }$-dressing method is based on the nonlocal $%
\overline{\partial }$-problem for a function with some normalization (see 
\textit{e.g.} [\ref{dbar}-\ref{kbook}]). We start with the following pair of 
$\overline{\partial }$-problems dual to each other 
\begin{equation}
\frac{\partial \,\chi ^{\prime }(\lambda ,\mu )}{\partial \,\overline{%
\lambda }}=\pi \,\delta (\lambda -\mu )+\iint_{\Bbb{C}}d\nu \wedge d%
\overline{\nu }\,\,\chi ^{\prime }(\nu ,\mu )\,\,R^{\prime }(\nu ,\lambda )
\label{e1}
\end{equation}
and 
\begin{equation}
\frac{\partial \,\chi ^{*}(\lambda ,\rho )}{\partial \,\overline{\lambda }}%
=-\pi \,\delta (\lambda -\rho )-\iint_{\Bbb{C}}d\nu \wedge d\overline{\nu }%
\,\,R(\lambda ,\nu )\,\,\chi ^{*}(\nu ,\rho )  \label{e2}
\end{equation}
where $\lambda \in \Bbb{C}$, bar means complex conjugation, $\delta (\lambda
)$ is the Dirac delta-function. The functions $\chi $, $\chi ^{*}$, $R$ and $%
R^{*}$ depend both on $\lambda $ and $\overline{\lambda }$, $\mu $ and $%
\overline{\mu }$ etc. To simplify the notations we will omit the dependence
on $\overline{\lambda }$, $\overline{\mu }$, $\overline{\rho }$, $\overline{%
\nu }$ etc. At $\lambda \rightarrow \mu $ we have 
\[
\chi ^{\prime }(\lambda ,\mu )=\frac{1}{\lambda -\mu }+\chi _{r}^{\prime
}(\lambda ,\mu )\quad ,\quad \chi ^{*}(\lambda ,\mu )=-\,\,\frac{1}{\lambda
-\mu }+\chi _{r}^{*}(\lambda ,\mu )
\]
where $\chi _{r}^{\prime }$ and $\chi _{r}^{*}$ are regular functions.
Solutions of the $\overline{\partial }$-problem with such properties have
been introduced in different contexts in [\ref{bogman},\ref{grinevich}]. We
shall refer to $\chi (\lambda ,\mu )$ as the Cauchy-Baker-Akhiezer (CBA)
functions. Further, we assume that $R^{\prime }(\nu ,\lambda )=R(\nu
,\lambda )=0$ for $\nu \in G$, $\lambda \in G$ where $G$ is certain domain
in $\Bbb{C}$ and $\mu ,\rho \in G$. So the functions $\chi _{r}(\lambda ,\mu
)$ and $\chi _{r}^{*}(\lambda ,\mu )$ are analytic in $G$ with respect to
both variables. Typically $G=D_{0}$ or $G=D_{0}\cup D_{\infty }$ where $D_{0}
$ and $D_{\infty }$ are the unit disks around the origin $\lambda =0$ and
around the infinity $\lambda =\infty $ , respectively. In general, $\chi $
and $R$ in (1),(2) are matrix valued functions.

To derive desired bilinear relations we first multiply from the right both
the sides of equation (1) by $f_{1}(\lambda )\,\chi ^{*}(\lambda ,\rho )$
and then multiply both the sides of equation (2) by $\chi ^{^{\prime
}}(\lambda ,\mu )$\thinspace $f_{2}(\lambda )\,$ from the left where $%
f_{1}(\lambda )$ and $f_{2}(\lambda )$ are arbitrary matrix-valued
functions. Summing up the obtained equations, one gets 
\[
\frac{\partial \,\chi ^{^{\prime }}(\lambda ,\mu )}{\partial \,\overline{%
\lambda }}\,f_{1}(\lambda )\,\chi ^{*}(\lambda ,\rho )+\chi ^{^{\prime
}}(\lambda ,\mu )\,\,f_{2}(\lambda )\,\frac{\partial \,\chi ^{*}(\lambda
,\rho )}{\partial \,\overline{\lambda }}=\,\,\quad \quad \quad \quad \qquad
\qquad \qquad \qquad \qquad \qquad \, 
\]
\[
\quad =\pi \,\delta (\lambda -\mu )\,f_{1}(\lambda )\,\chi ^{*}(\lambda
,\rho )-\pi \,\delta (\lambda -\rho )\,\,\chi ^{^{\prime }}(\lambda ,\mu
)\,\,f_{2}(\lambda )\,+\qquad \qquad \qquad \qquad \quad 
\]
\begin{equation}
+\iint_{\Bbb{C}}d\nu \wedge d\overline{\nu }\,\left[ \,\chi ^{^{\prime
}}(\nu ,\mu )\,R^{^{\prime }}(\nu ,\lambda )\,f_{1}(\lambda )\,\chi
^{*}(\lambda ,\rho )-\chi ^{^{\prime }}(\lambda ,\mu )\,f_{2}(\lambda
)\,R(\lambda ,\nu )\,\,\chi ^{*}(\nu ,\rho )\right] \,\,.  \label{e3}
\end{equation}
Integrating (\ref{e3}) with respect $\lambda $ over $\Bbb{C}$, one gets 
\[
\iint_{\Bbb{C}}d\lambda \wedge d\overline{\lambda }\,\,\left[ \frac{\partial
\,\chi ^{^{\prime }}(\lambda ,\mu )}{\partial \,\overline{\lambda }}%
\,f_{1}(\lambda )\,\chi ^{*}(\lambda ,\rho )+\chi ^{^{\prime }}(\lambda ,\mu
)\,\,f_{2}(\lambda )\,\frac{\partial \,\chi ^{*}(\lambda ,\rho )}{\partial \,%
\overline{\lambda }}\right] =\qquad \quad \qquad \qquad 
\]
\[
=2\pi \,i\,[\chi ^{^{\prime }}(\rho ,\mu )\,f_{2}(\rho )-f_{1}(\mu )\,\chi
^{*}(\mu ,\rho )]+\qquad \qquad \qquad \qquad \qquad \qquad \quad 
\]
\begin{equation}
+\iint_{\Bbb{C}}d\lambda \wedge d\overline{\lambda }\,\iint_{\Bbb{C}}d\nu
\wedge d\overline{\nu }\,\,\chi ^{^{\prime }}(\nu ,\mu )\,\left[ R^{^{\prime
}}(\nu ,\lambda )\,f_{1}(\lambda )-f_{2}(\nu )\,R(\nu ,\lambda )\,\right]
\,\chi ^{*}(\lambda ,\rho )\,.  \label{e4}
\end{equation}
Then integration of (\ref{e3}) over $\Bbb{C}/G$ gives 
\[
\iint_{\Bbb{C}/G}d\lambda \wedge d\overline{\lambda }\,\,\,\left[ \frac{%
\partial \,\chi ^{^{\prime }}(\lambda ,\mu )}{\partial \,\overline{\lambda }}%
\,f_{1}(\lambda )\,\chi ^{*}(\lambda ,\rho )+\chi ^{^{\prime }}(\lambda ,\mu
)\,\,f_{2}(\lambda )\,\frac{\partial \,\chi ^{*}(\lambda ,\rho )}{\partial \,%
\overline{\lambda }}\right] =\qquad 
\]
\begin{equation}
=\iint_{\Bbb{C}/G}d\lambda \wedge d\overline{\lambda }\,\iint_{\Bbb{C}%
/G}d\nu \wedge d\overline{\nu }\,\,\chi ^{^{\prime }}(\lambda ,\mu
)\,\,\left[ R^{^{\prime }}(\nu ,\lambda )\,f_{1}(\lambda )-f_{2}(\nu
)\,R(\nu ,\lambda )\,\right] \,\chi ^{*}(\lambda ,\rho )\,.  \label{e5}
\end{equation}

Considering equation (5) with $R^{\prime }=R$ (hence, $\chi ^{\prime }=\chi $%
) and $f_{1}=f_{2}=1$, one readily gets the well-known result $\chi ^{*}(\mu
,\rho )=\chi (\rho ,\mu )$ [\ref{bogkon},\ref{zman}].

The bilinear identities (4) and (5) (with $\chi ^{*}(\lambda ,\rho )=\chi
(\rho ,\lambda )$) are the fundamental bilinear relations within the $%
\overline{\partial }$-dressing method. We shall show that these relations
provide us integrable hierarchies and basic formulae associated with them in
a simple and transparent way.

In what follows we will consider the particular case of $f_{1}(\lambda
)=f_{2}(\lambda )=f(\lambda )$ and $\frac{\partial \,f(\lambda )}{\partial \,%
\overline{\lambda }}=0$ at $\lambda \in \Bbb{C}/G$ and assume that $%
f(\lambda )$ and $\chi (\lambda ,\mu )$ have no discontinuities on $\partial
G$. Thus, our starting bilinear relations are 
\[
2\pi \,i\,[f(\mu )\,\chi (\rho ,\mu )-\chi ^{^{\prime }}(\rho ,\mu )\,f(\rho
)]=-\iint_{\Bbb{C}}d\lambda \wedge d\overline{\lambda }\,\,\chi ^{^{\prime
}}(\lambda ,\mu )\,\,\frac{\partial \,f(\lambda )}{\partial \,\overline{%
\lambda }}\,\,\chi (\rho ,\lambda )+\quad 
\]
\begin{equation}
-\iint_{\Bbb{C}}d\lambda \wedge d\overline{\lambda }\,\iint_{\Bbb{C}}d\nu
\wedge d\overline{\nu }\,\,\,\chi ^{^{\prime }}(\nu ,\mu )\,\left[
R^{^{\prime }}(\nu ,\lambda )\,f(\lambda )-f(\nu )\,R(\nu ,\lambda )\right]
\chi (\rho ,\lambda )\quad ,  \label{e6}
\end{equation}
\[
\int_{\partial G}d\lambda \,\,\chi ^{^{\prime }}(\lambda ,\mu )\,f(\lambda
)\,\chi (\rho ,\lambda )=\qquad \qquad \qquad \qquad \qquad \qquad \qquad
\qquad \quad \quad 
\]
\begin{equation}
=\iint_{\Bbb{C}/G}d\lambda \wedge d\overline{\lambda }\,\iint_{\Bbb{C}%
/G}d\nu \wedge d\overline{\nu }\,\,\chi ^{^{\prime }}(\nu ,\mu )\,\left[
R^{^{\prime }}(\nu ,\lambda )\,f(\lambda )-f(\nu )\,R(\nu ,\lambda )\right]
\chi (\rho ,\lambda )\quad .  \label{eq7}
\end{equation}
At $f=1$ the relation (\ref{eq7}) gives 
\[
\chi ^{^{\prime }}(\rho ,\mu )-\chi (\rho ,\mu )=\qquad \qquad \qquad \qquad
\qquad \qquad \qquad \qquad \qquad \qquad \qquad 
\]
\begin{equation}
=-\frac{1}{2\pi \,i}\iint_{\Bbb{C}/G}d\lambda \wedge d\overline{\lambda }%
\,\iint_{\Bbb{C}/G}d\nu \wedge d\overline{\nu }\chi ^{^{\prime }}(\nu ,\mu
)\,\left[ R^{^{\prime }}(\nu ,\lambda )\,-\,R(\nu ,\lambda )\right] \chi
(\rho ,\lambda )\,\,.  \label{e8}
\end{equation}
Thus, in particular, 
\begin{equation}
\frac{\delta \,\chi (\rho ,\mu )}{\delta \,R(\nu ,\lambda )}=-\frac{1}{2\pi
\,i}\,\,\chi (\rho ,\lambda )\,\,\chi (\nu ,\mu )\qquad ,\qquad \rho ,\mu
\in G\,\,,\quad \,\nu ,\lambda \in \Bbb{C}/G\quad .  \label{e9}
\end{equation}
Then in the case of general degenerate variation of $R$ the formula (\ref{e8}%
) provides us an explicit transformation of $\chi $. Indeed, let 
\begin{equation}
R^{^{\prime }}(\nu ,\lambda )=R^{^{\prime }}(\nu ,\lambda )-2\pi
\,i\sum_{k=1}^{n}A_{k}(\nu )\,B_{k}(\lambda )  \label{e10}
\end{equation}
where $A_{k}$ and $B_{k}$ are arbitrary functions. Substituting (\ref{e9})
into (\ref{e8}), one gets 
\begin{equation}
\chi ^{^{\prime }}(\rho ,\mu )-\chi (\rho ,\mu
)=\sum_{k=1}^{n}X_{k}^{*^{\,\,\prime }}(\mu )\,X_{k}(\rho )  \label{e11}
\end{equation}
where 
\begin{equation}
X_{k}^{*^{\,\,\prime }}(\mu )\,=\iint_{\Bbb{C}/G}d\nu \wedge d\overline{\nu }%
\,\,\chi ^{^{\prime }}(\nu ,\mu )\,\,A_{k}(\nu )\quad ,\quad X_{k}(\rho
)\,=\iint_{\Bbb{C}/G}d\lambda \wedge d\overline{\lambda }\,\,B_{k}(\lambda
)\,\chi (\rho ,\lambda )\quad .  \label{e12}
\end{equation}
It follows from (\ref{e11}) that 
\begin{equation}
X_{i}^{*\,\,^{\prime }}(\mu )-X_{i}(\mu )=\sum_{k=1}^{n}X_{k}^{*^{\,\,\prime
}}(\mu )\,C_{k\,i}  \label{e13}
\end{equation}
where 
\begin{equation}
C_{k\,i}=\iint_{\Bbb{C}/G}d\lambda \wedge d\overline{\lambda }\,\,\iint_{%
\Bbb{C}/G}d\nu \wedge d\overline{\nu }\,\,B_{k}(\nu )\,\,\chi (\lambda ,\nu
)\,\,A_{i}(\lambda )\qquad .  \label{e14}
\end{equation}
Using (\ref{e13}) and (\ref{e11}), one gets 
\begin{equation}
\chi ^{^{\prime }}(\rho ,\mu )=\chi (\rho ,\mu )+\sum_{i,k=1}^{n}X_{i}(\mu
)\,\left[ (1-C)^{-1}\right] _{i\,k}\,\,X_{k}(\rho )  \label{e15}
\end{equation}
where $X_{i}(\lambda )$ are given by (\ref{e12}). This formula describes
dressing of the CBA function $\chi (\lambda ,\mu )$ under generic degenerate
transformation (\ref{e10}) of the $\overline{\partial }$-kernel on arbitrary
background $R(\nu ,\lambda )$ . In the particular case of degenerate
background kernel $R(\nu ,\lambda )$ and within a different approach,
similar formula has been derived recently in [\ref{manas}].

Now let us consider continuous transformations. The simplest of them are
given by similarity transformation of the kernel $R$%
\begin{equation}
R^{^{\prime }}(\nu ,\mu )=G(\nu )\,R(\nu ,\lambda )\,G^{-1}(\lambda )
\label{e16}
\end{equation}
where $G(\lambda )$ is a matrix-valued function. We assume that $G(\lambda )$
is analytic in $\Bbb{C}/G$ and continuous on $\partial G$. Considering the
formulae (\ref{e6}) and (\ref{eq7}) with $f(\lambda )=G(\lambda )$, we
conclude that under the transformations (\ref{e16}) the following bilinear
relations hold 
\begin{equation}
\chi ^{^{\prime }}(\rho ,\mu )\,G(\rho )-G(\mu )\,\chi (\rho ,\mu )=-\frac{1%
}{2\pi \,i}\,\iint_{G}d\lambda \wedge d\overline{\lambda }\,\,\,\chi
^{^{\prime }}(\lambda ,\mu )\,\,\frac{\partial \,G(\lambda )}{\partial \,%
\overline{\lambda }}\,\,\chi (\rho ,\lambda )  \label{e17}
\end{equation}
and 
\begin{equation}
\int_{\partial G}d\lambda \,\,\chi ^{^{\prime }}(\lambda ,\mu
)\,\,\,G(\lambda )\,\,\chi (\rho ,\lambda )=0\quad .  \label{e18}
\end{equation}
It is easy to check that these two relations are equivalent to each other.

Representing $G(\lambda )$ as $G(\lambda )=g^{^{\prime }}(\lambda
)\,\,g^{-1}(\lambda )$ and denoting $\chi (\lambda ,\mu )\equiv \chi
(\lambda ,\mu ;g)$, $\chi ^{^{\prime }}(\lambda ,\mu )\equiv \chi (\lambda
,\mu ;g^{^{\prime }})$ , one rewrites (\ref{e18}) in the form 
\begin{equation}
\int_{\partial G}d\lambda \,\,\chi ^{^{\prime }}(\lambda ,\mu ;g^{^{\prime
}})\,g^{^{\prime }}(\lambda )\,g^{-1}(\lambda )\,\,\chi (\rho ,\lambda
;g)=0\quad ,  \label{e19}
\end{equation}
that is the generalized Hirota bilinear identity introduced and discussed in
[\ref{bog1},\ref{bogkon2}]. In the particular case $\mu =\rho =0$ it
represents itself the celebrated Hirota bilinear identity (see \textit{e.g.}
[\ref{jimbo}]). It was shown in [\ref{bogkon2}] that the identity (\ref{e19}%
) provides an effective tool to describe and analyze the so-called
generalized integrable hierarchies and hierarchies of corresponding
singularity manifold equations.

The formulae (\ref{e17}) and (\ref{e18}) define finite continuous
transformations. For infinitesimal transformations $G(\lambda
)=1+\varepsilon \,\omega (\lambda )$ , $\delta R(\lambda ,\mu )=\varepsilon
\,\frac{\partial \,R(\lambda ,\mu )}{\partial \,\tau }$ and $\delta \chi
(\lambda ,\mu )=\varepsilon \,\frac{\partial \,\chi (\lambda ,\mu )}{%
\partial \,\tau }$ where $\varepsilon \rightarrow 0$ and $\tau $ is the
parameter of transformation. The infinitesimal version of the formulae (\ref
{e16})-(\ref{e18}) looks like 
\begin{equation}
\frac{\partial }{\partial \,\tau }\,R(\nu ,\lambda )=\omega (\nu )\,R(\nu
,\lambda )-R(\nu ,\lambda )\,\omega (\lambda )\qquad \quad ,  \label{e20}
\end{equation}
\[
\frac{\partial }{\partial \,\tau }\,\chi (\rho ,\mu )=\omega (\mu )\,\,\chi
(\rho ,\mu )-\chi (\rho ,\mu )\,\,\omega (\rho )-\qquad \quad \quad 
\]
\begin{equation}
\quad -\frac{1}{2\pi \,i}\iint_{G}d\lambda \wedge d\overline{\lambda }%
\,\,\,\chi (\lambda ,\mu )\,\,\frac{\partial \,\omega (\lambda )}{\partial \,%
\overline{\lambda }}\,\,\chi (\rho ,\lambda )\quad \quad ,  \label{e21}
\end{equation}
\begin{equation}
\frac{\partial }{\partial \,\tau }\,\chi (\rho ,\mu )=\frac{1}{2\pi \,i}%
\int_{\partial G}d\lambda \,\,\,\chi (\lambda ,\mu )\,\,\,\omega (\lambda
)\,\,\chi (\rho ,\lambda )\quad \quad ,  \label{e22}
\end{equation}

The formula (\ref{e21}) and (\ref{e22}) are equivalent to each other but in
some cases one of them is more convenient than the other. The formula (\ref
{e22}) with $\varepsilon \,\omega (\lambda )=\delta g(\lambda
)\,g^{-1}(\lambda )$ can be found also in [\ref{bogkon2}] while a version of
the formula (\ref{e21}) with integration over $\Bbb{C}$ has been derived in [%
\ref{zman}] (see also [\ref{kbook}]). A formula similar to (\ref{e22}) has
been derived in [\ref{grinevich}] by different method.

Equations (\ref{e21}) and (\ref{e22}) define integrable deformations of CBA
function since the $\overline{\partial }$-problems (\ref{e1}) and (\ref{e2})
allow to construct wide classes of exact solutions for them. Concrete form
of these integrable evolutions is defined by a form of the function $\omega
(\lambda )$. In the rest of the paper we will consider only scalar
equations. With the simplest choice $\omega (\lambda )=\frac{1}{2\pi \,i}\,%
\frac{1}{\lambda -a}$ where $a\in G$ is a parameter, one gets ($\tau =a$)
for $a\neq \rho $, $a\neq \mu $%
\begin{equation}
\frac{\partial \,\chi (\rho ,\mu )}{\partial \,a}=\left( \frac{1}{\mu -a}-%
\frac{1}{\rho -\mu }\right) \,\chi (\rho ,\mu )+\chi (a,\mu )\,\chi (\rho
,a)\quad ,\quad \rho \neq \mu \,\,.  \label{e23}
\end{equation}
In terms of the function $\beta (\rho ,\mu )$ defined as $\beta (\rho ,\mu
,a)=-\frac{\rho \,(\mu -a)}{\mu \,(\rho -a)}$ $\chi (\rho ,\mu ,a)$ equation
(\ref{e23}) looks like 
\begin{equation}
\frac{\partial \,\beta (\rho ,\mu )}{\partial \,a}=\beta (a,\mu )\,\beta
(\rho ,a)\quad .  \label{e24}
\end{equation}
Equation (\ref{e23}) (or (\ref{e24})) describes integrable deformations of
the CBA function due to the motion of position $a$ of the pole of $\omega
(\lambda )$ (see also [\ref{bogkon}]). In addition to this analytic meaning,
it has a pure geometric interpretation. Namely, equation (\ref{e24})
together with its cyclic permutations is nothing but the continuous analog
of the Darboux system $\frac{\partial \,\beta _{ik}}{\partial \,X_{l}}=\beta
_{il}\,\beta _{lk}$ which describes the triply conjugate system of surfaces
in $\Bbb{R}^{3}$ [\ref{Darboux}]. This old geometric system and its discrete
generalizations have attracted considerably interest recently (see \textit{%
e.g.} [\ref{kbook}, \ref{bogkon}, \ref{doliwa}, \ref{manas}]). Note that in
our approach the continuous Darboux system (\ref{e24}) arises in a scalar
case. In different context such a fact has been already mentioned in [\ref
{bogkon2},\ref{konmart}].

The continuous Darboux system (\ref{e24}) possesses all properties of the
standard Darboux system. In particular, the functions $X_{i}$ and $X_{i}^{*}$
defined by the formula (\ref{e12}) represent themselves the tangent vectors,
while the function 
\begin{equation}
\phi =\iint_{\Bbb{C}/G}d\lambda \wedge d\overline{\lambda }\,\iint_{\Bbb{C}%
/G}d\mu \wedge d\overline{\mu }\,\,A(\mu )\,\,\chi (\lambda ,\mu
)\,\,B(\lambda )  \label{e25}
\end{equation}
is a position vector. The formula (\ref{e15}) gives explicit transformation
of solution of the continuous Darboux system (\ref{e24}). It has a form of
the standard Darboux-Levy' transformation (see \textit{e.g.} [\ref{eisenhart}%
]). The choice $\omega (\lambda )=\frac{1}{2\pi \,i}\,\sum_{k=1}^{n}\,\frac{1%
}{\lambda -a_{k}}$ in (\ref{e21}), (\ref{e22}) leads to the system of $n$
separated continuous Darboux systems.

If now we parametrize the function $g$ in (\ref{e19}) as $g(\lambda )=\exp
\left( \sum_{n=1}^{\infty }\,\frac{t^{n}}{\lambda ^{n}}\right) $ then we
have infinite set of infinitesimal shifts of variables $t_{n}$ with $\omega
_{n}=g_{t_{n}}\,g^{-1}=\frac{1}{\lambda ^{n}}$ and the corresponding
equations (\ref{e22}) take the form 
\[
\frac{\partial \,\chi (\rho ,\mu )}{\partial \,t_{n}}=\left( \frac{1}{\mu
^{n}}-\frac{1}{\rho ^{n}}\right) \chi (\rho ,\mu )+\frac{1}{(n-1)!}\left\{ 
\frac{\partial ^{n-1}}{\partial \lambda ^{n-1}}\,\left[ \chi (\lambda ,\mu
)\,\chi (\rho ,\lambda )\right] \right\} _{\lambda =0}
\]
\begin{equation}
\qquad \qquad \qquad \qquad \qquad \qquad \qquad \qquad \qquad \qquad \quad
\quad \quad n=1,2,3,...\,\,\qquad .  \label{e26}
\end{equation}
This hierarchy of equations is equivalent to that studied in [\ref{bogkon2}]
and hence the hierarchy (\ref{e26}) describes the generalized
Kadomtsev-Petviashvili (KP) hierarchy which include the KP hierarchy itself,
the modified KP hierarchy and the hierarchy of KP singularity manifold
equations.

It is known that the times $t_{n}$ and the pole type parametrizations of the
KP hierarchy are connected by the Miwa transformation $t_{n}=\frac{1}{n}%
\sum_{i=1}^{\infty }a_{i}^{n}$ [\ref{miwa}]. In fact, due to the relation $%
\frac{\partial }{\partial \,a}=\sum_{n=1}^{\infty }\,a^{n-1}\frac{\partial }{%
\partial \,t_{n}}$ , the equivalence of the infinite hierarchy (\ref{e26})
and equation (\ref{e23}) is an easy check (see also [\ref{konmart}]).

Special choice of the function $\omega (\lambda )$ may provide interesting
deformations. For example, let us put $\omega (\lambda )=S(\lambda )$ where $%
S(\lambda )\,$is the Schwarz function of the curve $\partial G$. The Schwarz
function completely characterize the curve and $\overline{\lambda }%
=S(\lambda )$ at $\lambda \in \partial G$ [\ref{davis}]. Thus for boundaries 
$\partial G$ such that $S(\lambda )$ is analytic outside $G$, one has
deformations 
\begin{equation}
\frac{\partial }{\partial \,\tau }\,\chi (\rho ,\mu )=-\int_{\partial
G}d\lambda \,\,\overline{\lambda }\,\,\chi (\lambda ,\mu )\,\,\chi (\rho
,\lambda )=-\int_{\partial G}d\lambda \,\,\chi (\lambda ,\mu )\,\,S(\lambda
)\,\chi (\rho ,\lambda )\quad .  \label{e27}
\end{equation}
Such deformations are defined by the form of the boundary $\partial G$ of
the domain $G$. If $G$ is the unit disc $D_{0}$ then $S(\lambda )=\frac{1}{%
\lambda }$ [\ref{davis}] and the deformation (\ref{e27}) is of the KP type (%
\ref{e26}). In the case when $G$ is a circle of the radius $1$ with the
centre at $\lambda _{0}$, then $S(\lambda )=\frac{1}{\lambda -\lambda _{0}}+%
\overline{\lambda }_{0}\,$and the deformation (\ref{e27}) ($\tau =\lambda
_{0}$) coincides with (\ref{e23}).

Not only continuous integrable equations but also discrete ones can be
easily derived from the basic bilinear equations (\ref{e6}), (\ref{eq7}).
For instance, treating the transformation (\ref{e16}) with $G(\lambda )=%
\frac{1}{\lambda -a}$ as the shift in the discrete variable $n$, namely, $%
R^{^{\prime }}(\nu ,\lambda ;n)=R(\nu ,\lambda ;n+1)=T_{a}\,R(\nu ,\lambda
;n)$ one readily gets from (\ref{eq7}) the equation 
\begin{equation}
\left( T_{a}-1\right) \,\psi (\rho ,\mu )=T_{a}\,\psi (a,\mu )\cdot \psi
(\rho ,a)\qquad ,\quad \rho \neq \mu  \label{e28}
\end{equation}
where 
\begin{equation}
\psi (\rho ,\mu ;n)=(\mu -\rho )\,\left( \frac{\mu -a}{\rho -a}\right)
^{n}\,\chi (\rho ,\mu ;n)  \label{e29}
\end{equation}
that is the discrete analog of the Darboux system (\ref{e24}). Discrete
Darboux system has been derived in [\ref{bogkon}] and then has been
intensively studied during the last years in the context of discrete
integrable nets (see \textit{e.g.} [\ref{doliwa}]).

The basic bilinear relations (\ref{e6}) and (\ref{eq7}) are useful also for
study of constraints of generic integrable hierarchies. Here we will show
how the scalar BKP and CKP hierarchies [\ref{jimbo}] arise within this
approach. For this purpose it is sufficient to use relations (\ref{e6}) and (%
\ref{eq7}) with $R^{^{\prime }}(\nu ,\lambda )=R(-\lambda ,-\nu )$ and
assume that the kernel $R$ satisfies the constraint 
\begin{equation}
R(-\lambda ,-\nu )\,\,F(\lambda )=F(\nu )\,\,R(\nu ,\lambda )  \label{e30}
\end{equation}
where $F(\lambda )$ is a function obeying the condition $F(-\lambda )=\pm
\,F(\lambda )$ . In this case the domain $G$ has to be symmetric under the
change $\lambda \rightarrow -\lambda $ . Such type of constraints in matrix
case have been discussed recently in [\ref{zman}] and [\ref{doliwa}].

First we note that a solution of the $\overline{\partial }$-problem (\ref{e1}%
) with the kernel $R^{^{\prime }}(\nu ,\lambda )=R(-\lambda ,-\nu )$ is
given by $\chi ^{^{\prime }}(\nu ,\lambda )=\chi (-\lambda ,-\nu )$%
\thinspace . Then, the relation (\ref{eq7}) with $f(\lambda )=F(\lambda )$
and the kernel $R$ which satisfies (\ref{e30}) takes the form 
\begin{equation}
\int_{\partial G}d\lambda \,\,\chi (-\mu ,-\lambda )\,\,F(\lambda )\,\,\chi
(\rho ,\lambda )=0\qquad .  \label{e31}
\end{equation}
As in generic case, we have the generalized Hirota identity (\ref{e19}) but
now the transformations (\ref{e16}) have to be compatible with the
constraint (\ref{e30}). This implies that $g^{-1}(\lambda )=g(-\lambda )$ .
Due to the constraint (\ref{e31}) the identity (\ref{e19}) (with $%
g^{-1}(\lambda )=g(-\lambda )$ ) can be rewritten in an equivalent forms.

First we consider the case $F=1$ . So, $R(-\lambda ,-\nu )=R(\nu ,\lambda )$
. Then constraint (\ref{e31}) implies that $\chi (-\mu ,-\rho )=\chi (\rho
,\mu )$ . Hence the generalized Hirota identity (\ref{e19}) looks like ($%
G=D_{0}$) 
\begin{equation}
\int_{\partial D_{0}}d\lambda \,\,\chi (\lambda ,\mu ;g^{^{\prime
}}\,)\,\,g^{^{\prime }}(\lambda )\,\,g(-\lambda )\,\,\chi (-\lambda ,-\rho
;g)=0\qquad .  \label{e32}
\end{equation}
At $\mu =\rho =0$, and with the parametrization of $g$ by standard KP times (%
$g(\lambda )=\exp \left[ \sum_{n=1}^{\infty }\frac{t_{2n-1}}{\lambda ^{2n-1}}%
\right] $ ), the relation (\ref{e32}) coincides with the Hirota bilinear
identity for scalar CKP hierarchy.

The treatment of the constraint (\ref{e30}) with $F=\frac{1}{\lambda }$ is a
little bit more envolved. First, the constraint (\ref{e31}) gives 
\begin{equation}
\frac{1}{\mu }\,\chi (\rho ,-\mu )+\frac{1}{\rho }\,\chi (\mu ,-\rho )=\chi
(\rho ,0)\,\,\chi (\mu ,0)\qquad .  \label{e33}
\end{equation}
Then the identity (\ref{e19}) with $g(\lambda )=\exp \left[
\sum_{n=1}^{\infty }\frac{t_{2n-1}}{\lambda ^{2n-1}}\right] $ and $%
t_{1}^{^{\prime }}=t_{1}+\varepsilon $ , $\varepsilon \rightarrow 0$ implies 
\begin{equation}
\int_{\partial D_{0}}d\lambda \,\,\left[ \left( \frac{\partial }{\partial
\,t_{1}}+\frac{1}{\lambda }\right) \,\,\chi (\lambda ,-\mu )\right] \,\cdot
\,\,\chi (\rho ,\lambda )=0\qquad .  \label{e34}
\end{equation}
Subtracting equation (\ref{e31}) with $F=\frac{1}{\lambda }$ from (\ref{e34}%
), one gets 
\begin{equation}
\int_{\partial D_{0}}d\lambda \,\,\left[ \left( \frac{\partial }{\partial
\,t_{1}}+\frac{1}{\lambda }\right) \,\,\chi (\lambda ,-\mu )-\frac{1}{%
\lambda }\,\,\chi (\mu ,-\lambda )\right] \,\cdot \,\,\chi (\rho ,\lambda
)=0\qquad .  \label{e35}
\end{equation}
For $\mu =0$ the quantity in the bracket in (\ref{e35}) has no singularities
in $D_{0}$. Hence, equation (\ref{e35}) implies that $\left( \frac{\partial 
}{\partial \,t_{1}}+\frac{1}{\rho }\right) \,\,\chi (\rho ,0)=\frac{1}{\rho }%
\,\,\chi (0,-\rho )$ or equivalently 
\begin{equation}
g^{-1}(\lambda )\,\,\chi (0,\lambda )=-\lambda \,\frac{\partial }{\partial
\,t_{1}}\,\left[ g(-\lambda )\,\,\chi (-\lambda ,0)\right] \qquad .
\label{e36}
\end{equation}
With the use of (\ref{e36}) one rewrites the Hirota identity (\ref{e19})
with $\mu =\rho =0$ in the form 
\[
\,\frac{\partial }{\partial \,t_{1}}\,\int_{\partial D_{0}}d\lambda
\,\,\lambda \,\,\chi (\lambda ,0;g^{^{\prime }})\,\,\,g^{^{\prime }}(\lambda
)\,\,g(-\lambda )\,\,\chi (-\lambda ,0;g)=0 
\]
and finally as 
\begin{equation}
\int_{\partial D_{0}}\frac{\lambda \,\,d\lambda \,}{2\pi \,i}\,\,\,\chi
(\lambda ,0;g^{^{\prime }})\,\,\,g^{^{\prime }}(\lambda )\,\,g(-\lambda
)\,\,\chi (-\lambda ,0;g)=-1\qquad .  \label{e37}
\end{equation}
This relation is just the Hirota bilinear identity for the scalar BKP
hierarchy (see [\ref{jimbo}]) written in terms of wavefunctions with
normalization $\frac{1}{\lambda }$ as $\lambda \rightarrow 0$ . In terms of
times $t_{2n-1}$ the equations of the BKP hierarchy are given by equations (%
\ref{e26}) with $n=2k-1$ , $k=1,2,3,...$ and $\mu =0$ or $\rho =0$ . It is a
straightforward check that the constraint (\ref{e33}) is compatible with
these equations.

In similar manner one can treat multicomponent KP hierarchies, Toda lattice
hierarchy and other type of constraints.

\smallskip

\smallskip \smallskip

\smallskip 
\noindent%
\textbf{Acknowledgments.} The author is grateful to L. Bogdanov and L.
Martinez Alonso for fruitful  discussions. This work is supported in part by
the Grant PRIN 97 ''\textit{Sintesi}''.

\smallskip

\smallskip \smallskip \smallskip \smallskip

\begin{center}
\textbf{References}
\end{center}

\begin{enumerate}
\item  \label{zmnp}Zakharov V.E., Manakov S.V., Novikov S.P. and Pitaevski
L.P., 1980, \textit{Theory of Solitons} (Moscow, Nauka).

\item  Ablowitz M.J. and Segur H., 1981, \textit{Solitons and the Inverse
Scattering Transform}, (Philadelphia, SIAM).

\item  \label{jimbo}Jimbo M. and Miwa T., 1983, Publ. RIMS Kyoto Univ., 
\textbf{19}, 943.

\item  \label{segal}Segal G. and Wilson G., 1985, Publ. Math. I.H.E.S., 
\textbf{61}, 5.

\item  \label{dbar}Zakharov V.E. and Manakov S.V., 1985, Funk. Anal. Pril.,%
\textbf{19}, 11.

\item  \label{bogman}Bogdanov L.V. and Manakov S.V., 1988, J. Phys. A: Math.
Gen., \textbf{21}, 4719.

\item  \label{kbook}Konopelchenko B.G.,1993, \textit{Solitons in
Multidimensions}, (Singapore, World Scientific).

\item  \label{bogkon}Bogdanov L.V. and Konopelchenko B.G.,1995, J. Phys. A:
Math. Gen., \textbf{28}, 4173.

\item  \label{zman}Zenchuk A.Z. and Manakov S.V., 1995, Teor. Mat. Fyz., 
\textbf{105}, 371.

\item  Zakharov V.E. and Manakov S.V., 1998, Doklady Math ., \textbf{57},
471.

\item  \label{doliwa}Doliwa A., Manakov S.V. and Santini P.M., 1998, Commun.
Math. Phys., \textbf{196}, 1.

\item  \label{grinevich}Grinevich P.G. and Orlov A.Yu., 1989, in \textit{%
Problems of Modern Quantum Field Theory}, (Belavin A.A. Ed.), p. 86.

\item  \label{manas}Manas M., Martinez Alonso L. and Medina E., J. Phys. A:
Math. Gen., (to be published).

\item  \label{bog1}Bogdanov L.V., 1995, Physica D, \textbf{87}, 58.

\item  \label{bogkon2}Bogdanov L.V. and B.G. Konopelchenko, 1998, J. Math.
Phys., \textbf{39}, 4683, 4701.

\item  \label{Darboux} Darboux G, 1910, \textit{Lecons sur les Systemes
Orthogonaux et les Coordonnes} \textit{Curvilignes}, (Paris, Hermann).

\item  \label{konmart}Konopelchenko B.G. and Martinez Alonso L., 1999, Phys.
Lett. A, \textbf{258}, 272.

\item  \label{eisenhart}Eisenhart L.P., 1923, \textit{Transformations of
Surfaces}, (Princeton, Princeton Univ. Press).

\item  \label{miwa}Miwa T., 1982, Proc. J. Acad. Ser.A, \textbf{58}, 9.

\item  \label{davis}Davis P.J., 1974,\textit{\ The Schwarz function and its
applications}, (Buffalo, MAA).
\end{enumerate}

\end{document}